\documentclass[proof]{WileyASNA-v1}

\articletype{Original Article}%

\received{-- -- 2018}
\revised{-- -- 2018}
\accepted{8 November 2018}

\begin{document}

\title{Optical precursors to X-ray binary outbursts}

\author[1]{D.M. Russell*}

\author[1]{D.M. Bramich}

\author[2,3]{F. Lewis}

\author[1]{A. AlMannaei}

\author[4]{T. Al Qaissieh}

\author[1,5]{A. Al Qasim}

\author[1]{A. Al Yazeedi}

\author[1,6]{M.C. Baglio}

\author[7,8,1]{F. Bernardini}




\author[9]{N. Elgalad}


\author[4]{A. Gabuya}


\author[10,11]{J.P. Lasota}




\author[4]{A. Palado}


\author[2]{P. Roche}

\author[9]{H. Shivkumar}


\author[1]{S. Udrescu}


\author[12,13,14]{G. Zhang}

\authormark{D. M. Russell \textsc{et al.}}

\address[1]{\orgname{New York University Abu Dhabi}, \orgaddress{\state{Abu Dhabi}, \country{UAE}}}

\address[2]{\orgdiv{Faulkes Telescope Project, School of Physics and Astronomy}, \orgname{Cardiff University}, \orgaddress{\state{Cardiff}, \country{UK}}}

\address[3]{\orgdiv{Astrophysics Research Institute}, \orgname{Liverpool John Moores University}, \orgaddress{\state{Liverpool}, \country{UK}}}

\address[4]{\orgname{Al Sadeem Observatory}, \orgaddress{\state{Abu Dhabi}, \country{UAE}}}

\address[5]{\orgdiv{Mullard Space Science Laboratory}, \orgname{University College London}, \orgaddress{\state{Dorking}, \country{UK}}}

\address[6]{\orgname{INAF--Osservatorio Astronomico di Brera}, \orgaddress{\state{Merate}, \country{Italy}}}

\address[7]{\orgname{INAF--Osservatorio Astronomico di Roma}, \orgaddress{\state{Roma}, \country{Italy}}}

\address[8]{\orgname{INAF--Osservatorio Astronomico di Capodimonte}, \orgaddress{\state{Napoli}, \country{Italy}}}




\address[9]{\orgname{Paris-Sorbonne University Abu Dhabi}, \orgaddress{\state{Abu Dhabi}, \country{UAE}}}


\address[10]{\orgdiv{Institut d'Astrophysique de Paris}, \orgname{CNRS et Sorbonne Universit\'e}, \orgaddress{\state{Paris}, \country{France}}}

\address[11]{\orgdiv{$^{11}$Nicolaus Copernicus Astronomical Center}, \orgname{Polish Academy of Sciences}, \orgaddress{\state{Warsaw}, \country{Poland}}}





\address[12]{\orgdiv{Yunnan Observatories}, \orgname{Chinese Academy of Sciences (CAS)}, \orgaddress{\state{Kunming}, \country{P.R. China}}}

\address[13]{\orgdiv{Key Laboratory for the Structure and Evolution of Celestial Objects}, \orgname{CAS}, \orgaddress{\state{Kunming}, \country{P.R. China}}}

\address[14]{\orgdiv{Center for Astronomical Mega-Science}, \orgname{CAS}, \orgaddress{\state{Beijing}, \country{P.R. China}}}


\corres{*D. M. Russell, New York University Abu Dhabi, PO Box 129188, Abu Dhabi, UAE. \email{dave.russell@nyu.edu}}


\abstract{Disc instability models predict that for X-ray binaries in quiescence, there should be a brightening of the optical flux prior to an X-ray outburst. Tracking the X-ray variations of X-ray binaries in quiescence is generally not possible, so optical monitoring provides the best means to measure the mass accretion rate variability between outbursts, and to identify the beginning stages of new outbursts. With our regular Faulkes Telescope/Las Cumbres Observatory (LCO) monitoring we are routinely detecting the optical rise of new X-ray binary outbursts before they are detected by X-ray all-sky monitors. We present examples of detections of an optical rise in X-ray binaries prior to X-ray detection. We also present initial optical monitoring of the new black hole transient MAXI J1820+070 (ASASSN-18ey) with the Faulkes, LCO telescopes and Al Sadeem Observatory in Abu Dhabi, UAE. Finally, we introduce our new real-time data analysis pipeline, the ``X-ray Binary New Early Warning System (XB-NEWS)'' which aims to detect and announce new X-ray binary outbursts within a day of first optical detection. This will allow us to trigger X-ray and multi-wavelength campaigns during the very early stages of outbursts, to constrain the outburst triggering mechanism.}

\keywords{X-rays: binaries -- accretion, accretion disks -- black hole physics -- stars: neutron}



\maketitle


\section{Introduction}\label{sec1}

X-ray transients appear when they brighten considerably -- their X-ray intensity increasing by a factor of up to $\sim$10$^8$ \citep[e.g.][]{chenet97}. The exact mechanism that triggers these X-ray outbursts remains uncertain, despite more than 50 years of observational studies \citep[for a timeline of key discoveries, see][]{shawch13}. These X-ray transients are low-mass X-ray binaries (LMXBs) -- binary systems containing a compact object -- either a black hole (BH) or a neutron star (NS), and a low-mass 
companion star that is filling its Roche lobe so that material falls towards the compact object.
In LMXBs, matter flows from the companion star onto the compact object via an accretion disc.
LMXBs in our Galaxy often lay dormant, spending years-to-decades in a state of quiescence (being faint, feeding at low accretion rates, with X-ray luminosities of $\sim$10$^{29}$--$10^{33.5}$ erg s$^{-1}$).
They are (largely) only discovered when they enter an outburst and are detected by X-ray all-sky monitoring satellites, which can typically detect sources above $\sim$10$^{35}$--$10^{36}$ erg s$^{-1}$. During these shorter periods of intense activity called outbursts, the X-ray emission is much higher (up to $\sim$10$^{38}$ erg s$^{-1}$) and may approach the Eddington luminosity limit, $L_{\rm Edd}$ \citep[e.g.][]{chenet97}.

In quiescence, the X-ray luminosity is so low that only the most sensitive, state of the art X-ray satellites such as \textit{Chandra}, \textit{XMM-Newton} and \textit{NuSTAR}, can observe them. Even when detected, the X-ray spectrum is not well characterised due to the low number of photons detected, but can be described by a simple power law with a photon index of $\sim$2 \citep[e.g.][]{plotet17}.
The X-ray emission is produced close to the compact object, towards the inner radius of the accretion disc, which is truncated. This cooler, fainter disc (compared to higher luminosities, during outbursts) cannot be detected at X-ray energies in quiescence; however, since its inner regions are cooler than during outburst, it emits predominantly at ultraviolet (UV) and optical wavelengths. In fact, a large fraction of quiescent LMXBs are easily detectable at optical wavelengths using moderate-size ground-based telescopes, even when they are in quiescence \citep[e.g.][]{zuriet03}.
Evidence for emission from the accretion disc (and the companion star) can be found using $\sim 0.4$-m to 4-m class optical telescopes. Given the sensitivity limits of X-ray telescopes, it is usually only possible to detect the initial stages of a new outburst by performing regular monitoring using optical telescopes.

\section{What causes outbursts?}\label{sec2}

The process of accretion is responsible for the extreme heat and luminosity in the disc.
Some aspects of the accretion process are fairly well understood \citep{franet02} but many pressing questions remain unanswered:

$\bullet$ What is the structure of the accretion flow in quiescence?

$\bullet$ Where, and how exactly are LMXB outbursts triggered?

$\bullet$ What \textit{causes} outbursts, and their recurrence times?

$\bullet$ Do outbursts start differently in BH and NS systems?

An outburst itself is initiated due to an instability in the accretion disc. In quiescence a cold disc fills up with matter until at some radius, its temperature reaches a critical value, ionizing hydrogen and triggering an outburst. This is described by the thermal-viscous disc instability model \citep[DIM;][]{lasoet01}. Heating fronts propagate through the disc until it is in a hot, bright state, reaching X-ray luminosities approaching $L_{\rm Edd}$. The DIM, modified to include irradiation of the disc by X-rays from the inner regions, can broadly explain the LMXB outburst cycle \citep*{coriet12} only if the inner disc is truncated during quiescence \citep{dubuet01}.
However, exactly where and when the mechanism responsible for triggering an outburst is operating remains elusive, due to the lack of data during the initial rise stages.

The DIM model predicts that for an LMXB outburst, the instability is triggered at some radius within the inner part of the accretion disc, and propagates outwards. This is coined an `inside-out' outburst \citep{smak84,ludwet98,lasoet01}; it does not start exactly at the inner disc edge, and heating fronts propagate both ways. In dwarf nova outbursts (accreting white dwarfs), and suggested in a few LMXB outbursts is an alternative triggering process -- an `outside-in' outburst \citep[e.g.][]{warnet03,shahet98}. This is triggered by a thermal instability in the outer disc which creates a heating front that propagates inwards to smaller radii.

In both inside-out and outside-in outbursts, the heating fronts propagate with a speed $\alpha c_{\rm s}$, where $c_{\rm s}$ is the sound speed. Eventually, the inner disc fills in on the viscous timescale, resulting in X-rays from the hot disc -- this predicts a delay of several days in the rise to outburst of the X-ray emission \textit{from the disc} with respect to the optical emission \citep[see discussion in][]{hameet9797,dubuet01,bernet16}. However, during the decay of outbursts, near quiescence the disc is considered to be truncated, and the X-rays originate in the hot inner flow. To compare to this, it is not clear at what stage the disc fills in during the rise of an outburst. The movement of matter through the radiatively inefficient inner flow is rapid compared to the viscous timescales of the disc filling in, so in the case of a truncated disc, we may expect a shorter X-ray delay (less than a few days) if the X-rays originate in this inner flow.
Optical and X-ray detections are needed at the beginning stages of an LMXB outburst, in order to tell if outbursts are triggered inside-out or outside-in, and to measure how long it takes (and to what extent) for the inner disc to fill in during the initial brightening. Only if this can be tested on a number of LMXB outbursts, will we truly understand what triggers these outbursts.

It is notoriously difficult to detect the initial stages of a new LMXB outburst, due to the sporadic nature of their start times and the lack of regular monitoring. 
Nowadays, X-ray all-sky monitors still do not have the sensitivity to detect LMXBs during the initial rising stages. The limiting sensitivities of the X-ray monitors such as \textit{MAXI} \citep[The Monitor of All-sky X-ray Image;][]{matset09} and \textit{BAT} (Burst Alert Telescope) on \textit{Swift} \citep{krimet13} are several orders of magnitude greater than the quiescent fluxes of most Galactic LMXBs.
If it is possible to obtain both optical and X-ray monitoring during the very early stages of an outburst, we will be in a position to answer the question ``\textit{Where, and how exactly are LMXB outbursts triggered?}''

\section{LMXB Optical Monitoring}\label{sec3}

Currently, the only way that regular monitoring can be used to detect the start of new outbursts is with optical telescopes. Queue-scheduled facilities are needed to monitor sources continuously, and robotic telescopes provide the best set-up for this purpose because observations can be performed remotely and do not require real-time human interaction. A few outburst rises have been detected at optical wavelengths before X-ray detection, and claims have been made from some of these that the optical rise preceded the X-ray rise \citep[e.g.][]{oroset97,buxtba04,bernet16}, but the initial rise of the X-ray emission was never detected because they were fainter than the detection limit of the X-ray instruments during the initial brightening, in all cases. X-ray 
observations have never been triggered 
quick enough for the early X-ray rise out of quiescence to be seen.

Many LMXBs can be detected with small optical telescopes in only a few minutes of exposure time.
We have been monitoring $\sim$40 LMXBs for $>10$ years using the 2-m Faulkes Telescope North (Maui, Hawaii, USA) and 2-m Faulkes Telescope South (Siding Spring, Australia)\footnote{\url{http://www.faulkes-telescope.com/xrb/}} \citep{lewi18}. 
The Faulkes Telescopes\footnote{\url{http://www.faulkes-telescope.com/}} are the largest telescopes in the Las Cumbres Observatory (LCO)\footnote{\url{https://lco.global/}}, a global robotic telescope network which also comprises a suite of 1-m and 0.4-m telescopes distributed over six continents \citep{browet13}.
We typically observe each LMXB once per week when they are visible (above the horizon at night) in three filters: V, R and i$^{\prime}$. The main aims of our monitoring campaign are to characterise the quiescent variability of LMXBs, gather multi-waveband light curves of outbursts to be included in multi-wavelength campaigns, and to detect new outbursts \citep{lewiet08}.
In addition, in the last few years the LCO 1-m network has become available to us. The 1-m network currently comprises of nine identical 1-metre telescopes, mostly in the southern hemisphere, making it possible to perform high cadence monitoring of sources in outburst, e.g. GS 1354--64 \citep{koljet16}.

\begin{figure*}[t]
	\centerline{\includegraphics[width=60mm,angle=270]{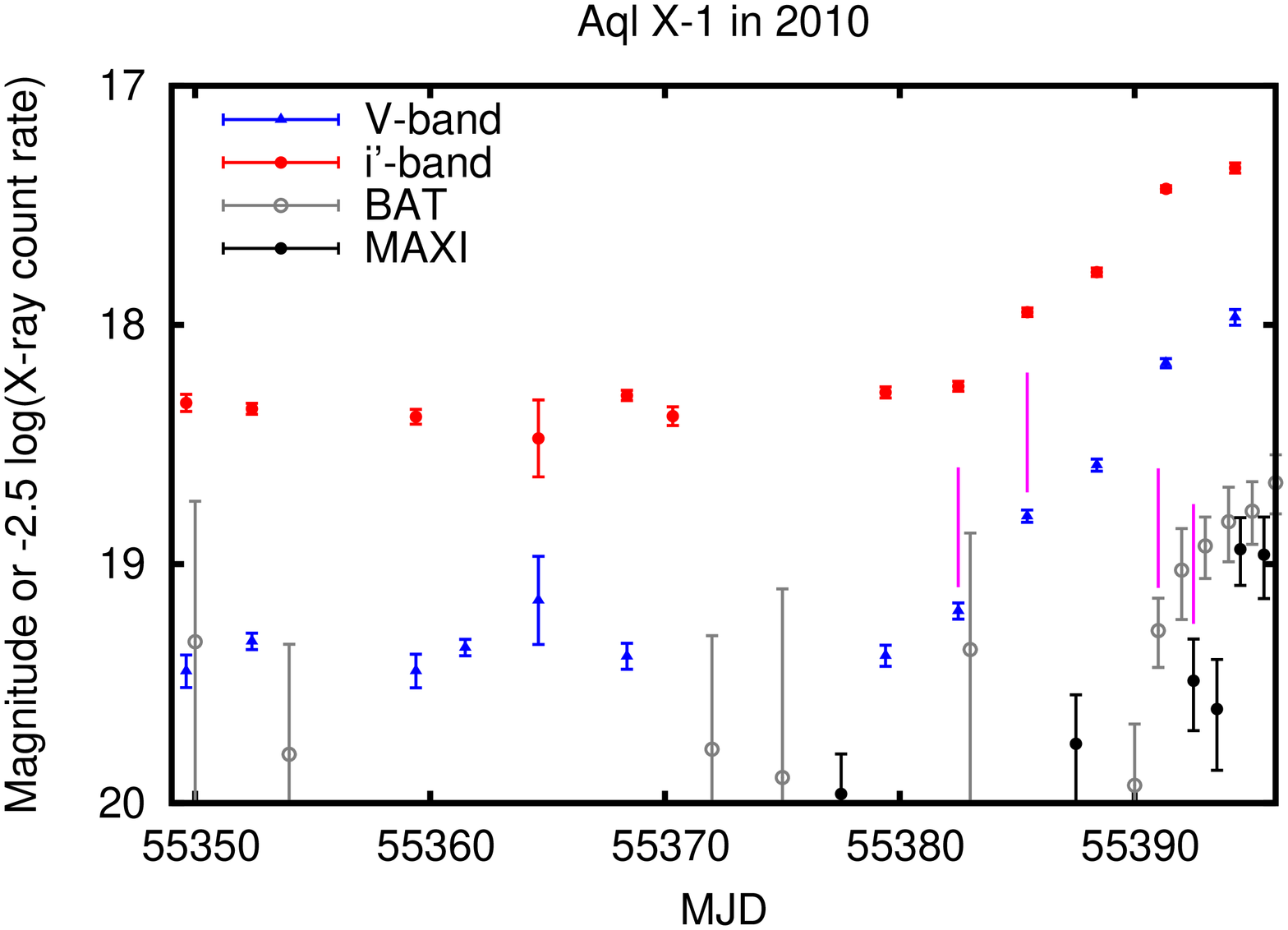}\includegraphics[width=60mm,angle=270]{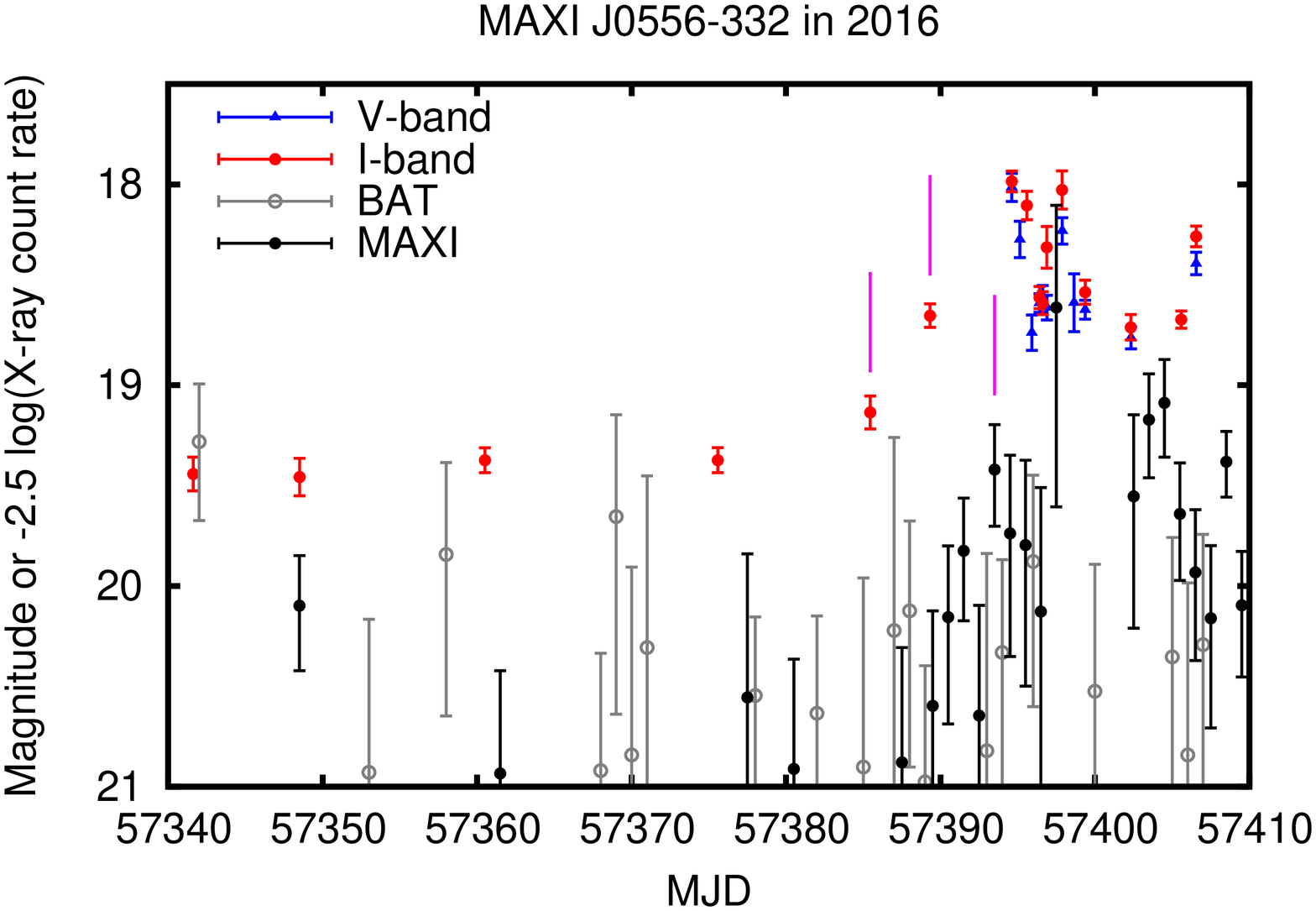}}
	\centerline{\includegraphics[width=60mm,angle=270]{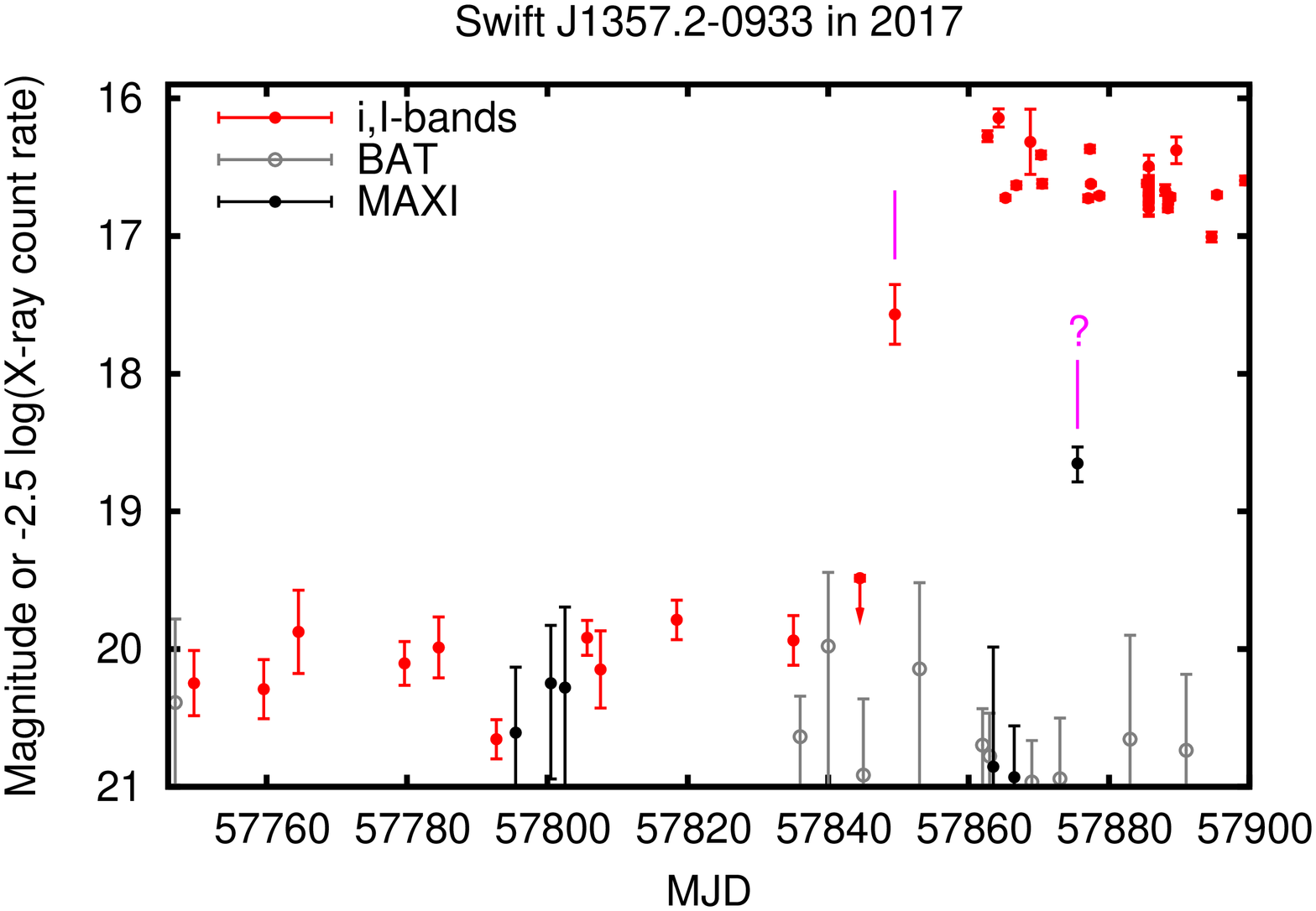}\includegraphics[width=60mm,angle=270]{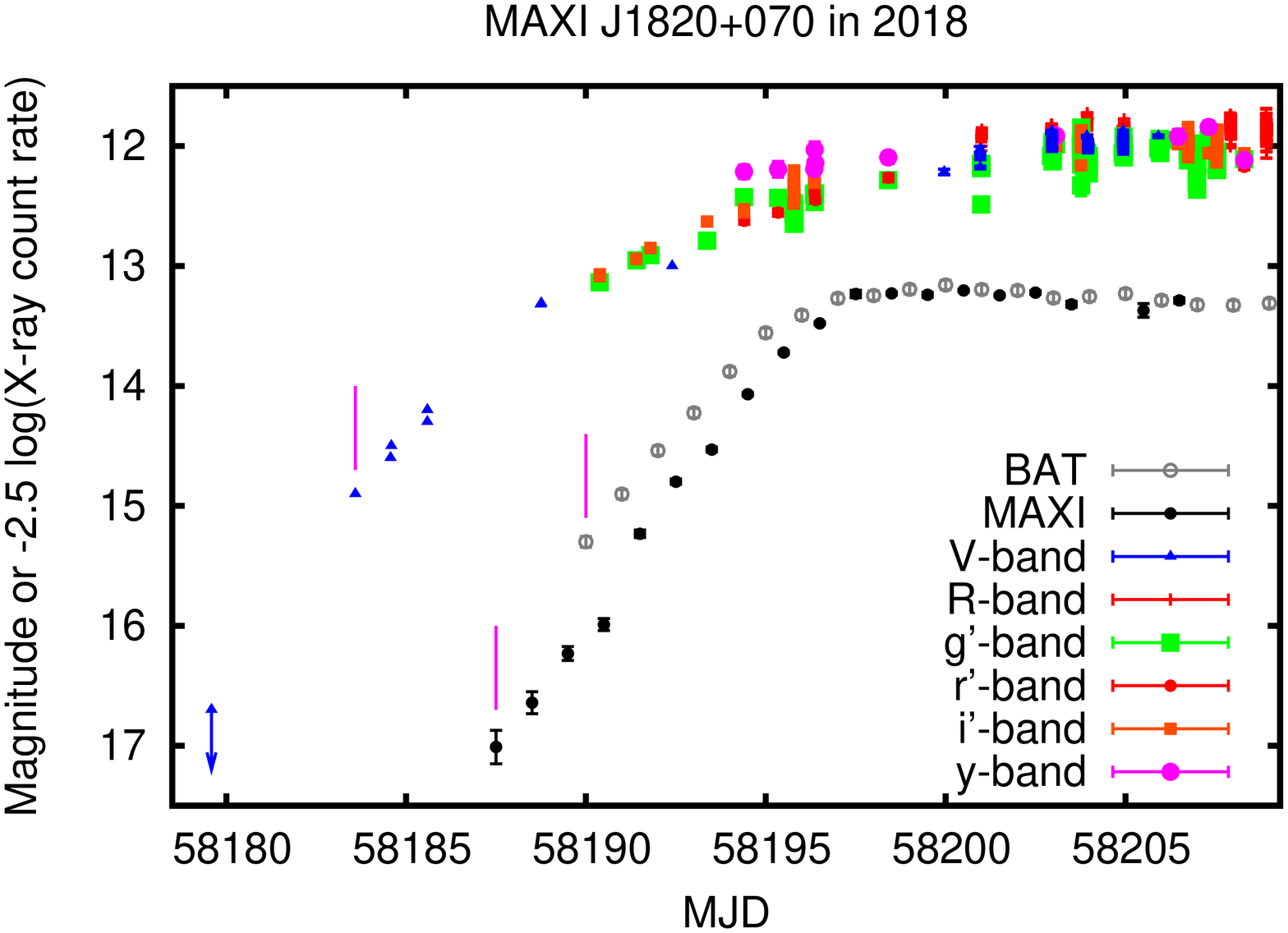}}
	\caption{Examples of LMXB outbursts detected by optical telescopes before the first X-ray detections. Soft X-ray (2--20 keV) data from \textit{MAXI} are shown in black and hard X-ray data (15--50 keV) from \textit{BAT} on \textit{Swift} are in grey. The purple vertical lines mark the dates of first significant detections at optical, then X-ray wavelengths. All were detected first with the Faulkes Telescopes except MAXI J1820+070, which was discovered as the optical transient ASASSN-18ey by the ASAS-SN survey.\label{fig-rises}}
\end{figure*}

Around 30--50\% of the LMXBs in our monitoring program are bright enough to be regularly detected in quiescence. However, we have not been able to detect most new outbursts as soon as they occur, because we have not had a real-time data analysis pipeline. The nature of issues with the data, and their evolution over the years, has generally required manual inspection of the data. As such, the data from only some of the sources, typically the more active ones, have been published to date. This has unfortunately resulted in missing a number of outburst brightenings, even though the data were taken.

We typically detect a new outburst a few days to a few weeks before the first X-ray detection by an X-ray all-sky monitor.
However, only in a few cases \citep[e.g.][]{lewiet08ATel1726,russle09ATel1970,lewiet12ATel4162,alqaet17ATel10075}
were we able to report the detection of the new outburst in order to trigger multi-wavelength follow-up observations.
Clearly, for this to occur regularly, an automated system is required. Based on the detection of an outburst from our Faulkes optical monitoring before the first X-ray detection in 15 of 17 outbursts, we estimate that with such a system set up, we will detect the initial optical rise of $\sim 80$--90\% of all outbursts from sources that we are monitoring (that are visible at the time of initial brightening). Having the light curves of all $\sim$40 LMXBs (including quiescent sources that we do not usually detect, but for which we would if an outburst occurred) updated automatically, daily, will allow us to detect the early stages of outbursts at multiple wavelengths.


\section{Examples of optical rises before X-ray detections}\label{sec4}

In Fig. \ref{fig-rises} we show the optical and daily X-ray all-sky monitor \citep[\textit{MAXI} and \textit{BAT} on \textit{Swift};][]{matset09,krimet13} light curves of the initial stages of four LMXB outbursts. The purple bars indicate the epochs at which the rise into outburst was first detected at optical and X-ray (3$\sigma$ detections) wavelengths. For the NS systems Aql X--1 and MAXI J0556--332 the optical rise of the outbursts were detected by our Faulkes monitoring about a week before the X-ray flux brightened to sufficient fluxes to be detected by the all-sky monitors \citep{russet10ATel2871,russet16ATel8517,russet16ATel8854}. For the BH LMXB Swift J1357.2--0933 \citep{russet18} the X-ray all-sky monitors did not detect the outburst at all (except one single \textit{MAXI} point 26 days after the first optical detection).

The new X-ray transient, MAXI J1820+070 (Fig. \ref{fig-rises}; lower right panel) was discovered in March 2018 by \textit{MAXI} \citep{kawaet18}. However, it was soon realised that a new optical transient, ASASSN-18ey, discovered five days earlier by the All-Sky Automated Survey for SuperNovae (ASAS-SN), was in fact the same source \citep{deni18}. We have been monitoring this bright LMXB with the Faulkes Telescopes and the LCO 1-m network in $g^{\prime}$,$r^{\prime}$,$i^{\prime}$,$y$-band filters, and we used some of the initial data to infer that the system likely contains a BH, and not a NS \citep{baglet18ATel11418}. We have also been monitoring MAXI J1820+070 with a Meade LX850 16-inch (41-cm) telescope using Baader LRGB CCD filters (similar central wavelengths to $g^{\prime}$,$V$,$R$-bands) at Al Sadeem Observatory\footnote{\url{http://alsadeemastronomy.ae/}} (Owner/Co-founder Thabet Al Qaissieh, Director/Co-founder Alejandro Palado, Resident Astronomer Aldrin B. Gabuya), located in Al Wathba South, outside the city of Abu Dhabi in the United Arab Emirates \citep{russet18ATel11533}. The light curve of MAXI J1820+070 in Fig. \ref{fig-rises} represents one of the best sampled rises into outburst at optical wavelengths of a LMXB to date \citep[we also include some magnitudes published in ATels;][]{deni18,littet18,gandet18}, and shows that the X-ray flux started to flatten and decay slightly, before the optical reached peak flux. MAXI J1820+070 provides evidence that we are now entering an era in which \textit{previously unknown} LMXBs are being discovered at optical wavelengths before X-rays. However, again the initial rise out of quiescence was missed, this time at both optical and X-ray wavelengths.

The 
short-lived 2015 outburst of V404 Cyg was the brightest LMXB outburst seen in decades. We detected the initial brightening of the outburst from our Faulkes Telescope monitoring -- the optical precursor occurred a week before the first X-ray flare was detected \citep{bernet16}. Since the X-ray rise was very rapid, the week delay suggests that the disc may have heated up before the X-ray outburst began. The X-ray delay is consistent with the time needed to refill the inner region and hence move the inner edge of the disc inwards, allowing matter to reach the central BH, finally causing the X-ray brightening. This may be the case for some of the other outbursts in which the optical rise was detected before X-ray confirmation (Fig. \ref{fig-rises}), 
but for these we have no constraint on when the X-rays started rising out of quiescence. Even though we are detecting new outbursts regularly at their early stages, the question ``\textit{Does the optical really rise before the X-ray?}'' remains elusive due to the lack of early X-ray detections.








\section{Introducing XB-NEWS}\label{sec5}

In order to alert the community to an optical brightening of an X-ray binary as soon as possible after it has been captured in an observation, an automated pipeline and alert system is required, thereby removing human actions and reaction times. As images are acquired, we have to manually download them, perform the
photometric measurements, and then build the light curves. This process is labour-intensive and it depends on the availability of someone to do the work in a timely fashion. 

We have recently been awarded a grant to fund the development of the automated data pipeline and alert system outlined above. We are currently developing the X-ray Binary New Early Warning System (XB-NEWS) to achieve this and to process our archive of monitoring observations in a systematic fashion. There are two essential components to such a system. The first component is a data pipeline that takes as input a raw image and associated calibration data, and outputs a calibrated photometric measurement of the required target. The second component is an alert pipeline that analyses the complete target light curve including the new data point and decides if anomalous behaviour is occurring. The alert pipeline can then send an automated message to the XB-NEWS team.


The first component of the system is already relatively well-developed as of August 2018. The system interrogates the LCO data archive\footnote{\url{https://archive.lco.global/}} at regular intervals (up to once every few minutes) for a pre-defined list of targets. Any new science data (raw and reduced) that is available in the archive for the required targets, including master calibration files (e.g. master flat frames), are downloaded and integrated into our local archive. A data quality control step is implemented locally. Work is currently being done on automatically identifying poor reductions due to low quality master flat frames, and then redoing the flat fielding stage with a better quality master flat frame from close-in-time to the image to be calibrated. The target flux will be (re-)measured and appended to the relevant light curve. All of the light curves for the targets will be visible online at a dedicated webpage.

One of the key aspects of the data pipeline is that it is designed to be highly robust and fully automatic. The delay between an image being made available in the LCO archive and the photometry being appended to the light curve will be of the order of 1--10 minutes, limited principally by the cadence at which the LCO archive is interrogated by the pipeline. Hence the pipeline is in essence a near real-time pipeline. We will make the data pipeline publicly available via \textit{GitHub} along with full documentation since it has a more general application to LCO data management and photometry. We also intend to re-reduce all of our older data with XB-NEWS and integrate the results into our local archive.

The alert system is yet to be developed. However, it is envisioned that each time a new photometric data point arrives, the relevant light curve will be re-analysed for anomalous behaviour (e.g. gradual brightening, outburst, fading, possibly colour changes). If anomalous behaviour is detected, then the system will alert the XB-NEWS team and we will write an alert in the form of an ATel if the behaviour is worthy of multi-wavelength attention. This will allow rapid multi-wavelength follow-up observations of the target when appropriate. We intend to investigate and harness the power of machine learning techniques for the anomaly detection where possible. We expect the first XB-NEWS announcements to be made towards the end of 2018, or at the start of 2019.



When we detect a new outburst, we will also trigger other facilities. For example, the sensitivity of \textit{Swift}, triggered within 1--2 days of the initial optical brightening, will help shed light on the question of if the optical rises first, if outbursts are inside-out or outside-in, and how the inner disc is filled by matter during the initial rise. For additional X-ray coverage, we have informal agreements to alert \textit{ASTROSAT} \citep{singet14}, \textit{HXMT} \citep[Hard X-ray Modulation Telescope;][]{zhanet14}, \textit{NICER} \citep[Neutron star Interior Composition Explorer;][]{gendet12} and \textit{INTEGRAL} \citep[INTErnational Gamma-Ray Astrophysics Laboratory;][]{winket03} teams of new outbursts. We will also aim to gather radio, mm and infrared rapid follow-up observations based on new detections, to catch the outburst rise over multiple wavelengths.


In addition to detecting the initial stages of new outbursts, the real-time monitoring of LMXBs (and potentially other variable sources) will allow us to detect unusual behaviour. For example, in 2016 we discovered, by analysing new data of the BH system Swift J1753.5--0127, that it had suddenly faded dramatically after an 11-year-long outburst \citep{russet16ATel9708,alqaet16,zhanet18}. This fade was not noticed at other wavelengths; it was too faint for X-ray all-sky monitors. In the past we have detected state changes and strong variability in some LMXBs from our monitoring \citep[e.g.][]{lewiet10ATel2459,lewiet10AA,russet10ATel2997}.
We are also expanding our monitoring programs beyond LMXBs, to systems such as cataclysmic variables \citep[e.g.][]{zhanet17}.



\section*{Acknowledgments}

DMR and DMB acknowledge the support of the NYU Abu Dhabi Research Enhancement Fund under grant RE124. JPL acknowledges support by the National Science Centre, Poland grant 2015/19/B/ST9/01099 and by a grant from the French Space Agency CNES. The research reported in this publication was supported by Mohammed Bin Rashid Space Centre (MBRSC), Dubai, UAE, under Grant ID number 201701.SS.NYUAD. The Faulkes Telescope Project is an education partner of Las Cumbres Observatory. The Faulkes Telescopes are maintained and operated by LCO. This research has made use of the MAXI data provided by RIKEN, JAXA and the MAXI team. Swift/BAT transient monitor results provided by the Swift/BAT team.













\bibliography{precursors}%



\end{document}